# Log normal claim models with common shocks

Greg Taylor

School of Risk and Actuarial Studies
University of New South Wales
UNSW Sydney, NSW 2052
AUSTRALIA

October 2022

**Abstract.** This paper is concerned with modelling multiple claim arrays that are subject to one or more common shocks. It uses a structure that involves very general forms both idiosyncratic and common shock components of cell means. The dependencies between arrays, or between cells within an array, generated by the shocks are also of very general form. All of this appears in the prior literature, where the idiosyncratic and shock components are additive. This has created the awkwardness of unbalanced shocks. The present paper rectifies this by defining these components as multiplicative. Observations in individuals cells of claim arrays are assumed log normal (later log Tweedie) in order to accommodate the multiplicativity. Conveniently, the log normal case reduced parameter estimation to linear regression, yielding closed form solution of location parameters, and even of dispersion parameters in some cases.

## 1. Introduction

Common shock models were introduced to the actuarial literature by Lindskog and McNeil (2003). These provide a means creating simple models of data sets that contain dependencies between observations.

They have been used in a variety of actuarial settings. For example, in claim modelling, Meyers (2007) applied the concept within the context of collective risk theory. Other fields of application include capital modelling (Furman and Landsman, 2010) and mortality modelling (Alai, Landsman and Sherris, 2013, 2016).

Avanzi, Taylor, Vu and Wong (2016) applied a common shock model to claim triangles, creating dependencies between triangles. In this model, cells were assumed Tweedie distributed, and the authors generated a multivariate Tweedie distribution to describe the triangles, including the shock.

Avanzi, Taylor and Wong (2018) established a very general framework for the introduction of common shocks into claim triangles. This included dependencies between corresponding cells of different triangles, corresponding accident, development or calendar periods, or combinations of these; as well as more general dependencies.

Many models of claim triangles are multiplicative, especially those resembling the chain ladder, with multiplicative row and column effects. When an additive common shock is added, it can cause an awkward feature whereby the proportion of claim experience contributed by the shock can vary widely over a triangle.

This is the “unbalanced” feature discussed by Avanzi, Taylor, Vu and Wong (2021). Those authors increased balance by modifying the shock to assume different magnitudes in different development periods. Subsequently, Taylor and Vu (2022) produced “auto-balanced” models, in which the awkward imbalance is totally eliminated.

Essentially, the difficulties of balance arise from the assumed structure of cell means, which contain a **multiplicative** chain-ladder-like term supplemented **additively** by a common shock.

This can be simplified by subjecting the claim experience in each cell to a log transform, thus producing an additive model, and supplementing this with an additive shock.

This would usually mean that the individual cell observations are the exponential of some distribution with an asymptotically exponential tail, such as log normal. This will be suitable for some data sets, and not others. In any case, log normal claim models are no stranger to the actuarial literature. See, for example, Taylor (2000), Wüthrich and Merz (2008), Shi, Basu and Meyers (2012).

The objective of the present paper is to adopt this log normal or similar structure, but apply it within the very general common shock framework of Avanzi, Taylor and Wong (2018), mentioned above. The benefits of doing so are, first, that balance of the shock is easily achieved; second, that parameter estimation is reduced to linear regression in the log normal case; and, third, that even variance estimation can sometimes be achieved in closed form.

The paper is concerned with claim triangles, and the forecast of loss reserves based on them. It proceeds as follows. After some notational and mathematical preliminaries in Section 2, both additive and multiplicative common shocks are described in Section 3. The full detail of a log normal common shock model, including model structure, parameter estimation and forecast, is discussed in Section 4, and Section 5 notes how these models may be generalized from log normal to log Tweedie. Section 6 discusses the special cases of these models in which the idiosyncratic (non-common-shock) component of each cell assumes a multiplicative chain ladder form, and Section 7 provides a numerical example. Section 8 winds up with some concluding remarks.

# 2. Framework and notation

## 2.1. Notation

The general common shock framework of Avanzi, Taylor and Wong (2018) is adopted. A claim array $\mathcal{A}$ will be defined here as a 2-dimensional array of random variables $X_{ij} > 0$, indexed by integers $i,j$, with $1 \le i \le I^*, 1 \le j \le f(i,I) \le J$ for some fixed integers $I^*, J$ and integer-valued function $f(.,.)$. The symbol $I^*$ has been used here, rather than the more natural $I$, since the latter will be used later to denote an identity matrix. For any given pair $i,j$, the random variable $X_{ij}$ may or may not be present.

The subscripts $i,j$ typically index accident period (row) and development period (column) respectively, and the $X_{ij}$ represent observations on claims, commonly claim counts or amounts. In the special case $I^* = J$ and $\mathcal{A} = \{X_{ij}: 1 \le i \le I^*, 1 \le j \le I - i + 1\}$, the array reduces to the well known claim triangle.

Define $t = i + j - 1$, so that $t = 1,2,\ldots,I^* + J - 1$. Observations with common $t$ lie on the $t$-th diagonal of $\mathcal{A}$.

Subsequent sections will often involve the simultaneous consideration of multiple business segments, with one array for each segment. A segment could be a line of business.

It will be necessary in this case to consider a collection $\mathbb{A} = \{\mathcal{A}^{(n)}, n = 1,2,\ldots,N\}$ of claim arrays, where $\mathcal{A}^{(n)}$ denotes the array for segment $n$. It will be assumed that all $\mathcal{A}^{(n)}$ are congruent, i.e. are of the same dimensions $I^*, J$, and that they have missing observations in the same $i, j$ locations. Consequently, the order $|\mathcal{A}^{(n)}|$ of $\mathcal{A}^{(n)}$ is the same for all n, and will be denoted $|\mathcal{A}|$.

The $i, j$ observation of $\mathcal{A}^{(n)}$ will be denoted $X_{ij}^{(n)}$; the entire $i$-th row of $\mathcal{A}^{(n)}$ denoted $\mathcal{R}_i^{(n)}$; and the entire $j$-th column $\mathcal{C}_j^{(n)}$. It will sometimes be useful to represent the enter array $\mathcal{A}^{(n)}$ as a vector. This will be denoted by $X^{(n)}$, with components $X_{ij}^{(n)}$ arranged in some pre-defined order. The order might conveniently be dictionary order but, in general, it is arbitrary. It must, however, be the same for all $n$.

The vector random variables $X^{(n)}$ may be stacked to form a larger vector

$$X = \begin{bmatrix} X^{(1)} \\ X^{(2)} \\ \vdots \\ X^{(N)} \end{bmatrix} \tag{2.1}$$

It will be convenient to write $Y_{ij}^{(n)} = ln\, X_{ij}^{(n)}$, and to stack the $Y_{ij}^{(n)}$ into larger vectors $Y^{(n)}$ and $Y$ in the same way as was done for the $X_{ij}^{(n)}$. Now introduce the notation

$\mu_{ij}^{(n)} = E\left[Y_{ij}^{(n)}\right], \mu^{(n)} = E\left[Y^{(n)}\right], \mu = E[\mathrm{Y}]$, and $\Sigma^{(n)} = Var\left[Y^{(n)}\right], \Sigma = Var[\mathrm{Y}]$.

It will also be useful to consider diagonals of $\mathcal{A}^{(n)}$, where the $t$-th diagonal is defined as the subset $\left\{X_{ij}^{(n)} \in \mathcal{A}^{(n)} : i + j - 1 = t\right\}$, and represents claim observations from the $t$-th calendar period, $t = 1$ denoting the calendar period in which the first accident period falls. The entire $t$-th diagonal of $\mathcal{A}^{(n)}$ will be denoted $\mathcal{D}_t^{(n)}$.

Let $t_{max}$ be the largest value of $t$ occurring in $\mathcal{A}^{(n)}$. Estimates of $X_{ij}^{(n)}$ for $t > t_{max}$ will be forecasts. Let the set of $X_{ij}^{(n)}$ requiring forecast be denoted by $\mathcal{A}^{(n)*}$. This will correspond with some set of ordered pairs $(i, j)$ with $t = i + j - 1 > t_{max}$.

An identity matrix will be denoted by $I$, and a column vector with all components equal to unity denoted by 1. When the dimensions of these quantities are not immediately apparent from the context, they will appear as subscripts to the quantities themselves. For example, $I_N$ will denote the $N \times N$ identity matrix, and $1_N$ the $N$-dimensional unit vector.

## 2.2. Mathematical preliminaries

There will be occasion to use the **Kronecker matrix product**. The Kronecker product of the $m \times n$ matrix $A = [a_{ij}]$ and $p \times q$ matrix $B = [b_{ij}]$ is the $mp \times nq$ matrix

$$A \otimes B = \begin{bmatrix} a_{11}B & a_{12}B & \cdots \\ a_{21}B & a_{22}B & \\ \vdots & & \ddots \end{bmatrix}. \tag{2.2}$$

It may be shown that

$$(A \otimes B)^T = A^T \otimes B^T, \tag{2.3}$$

where the upper $T$ denotes matrix transposition.

It will also be useful to note the so-called **mixed-product property**, according to which

$$(A \otimes B)(C \otimes D) = AC \otimes BD, \tag{2.4}$$

where, here and below, vectors and matrices are assumed of suitable dimensions for the products and Kronecker products that appear.

Further, if $a$ is a column vector, then

$$(a \otimes B)C = a \otimes (BC), \tag{2.5}$$

and

$$C(a^T \otimes B) = a^T \otimes (CB). \tag{2.6}$$

## 2.3. Probabilistic preliminaries

### 2.3.1. Univariate Tweedie family

The Tweedie family is a sub-family of the **exponential dispersion family**. The latter has two well known representations, the additive and reproductive forms (Jorgensen, 1987). In common with Avanzi, Taylor, Vu and Wong (2016), the present paper develops the multivariate Tweedie family (Section 2.3.2) within the context of the **additive representation**, which has the pdf

$$p(X = x; \theta, \lambda) = a(x, \lambda) exp\{x\theta - \lambda b(\theta)\}, \tag{2.7}$$

where $\theta$ is a location parameter, $\lambda > 0$ is a dispersion parameter, and $b(\theta)$ is called the cumulant function.

This distribution has cumulant generating function

$$K_X(t) = \lambda[b(\theta + t) - b(\theta)], \tag{2.8}$$

giving

$$E[X] = \lambda b'(\theta), \tag{2.9}$$

$$Var[X] = \lambda b''(\theta). \tag{2.10}$$

The **Tweedie sub-family** is obtained by the selection

$$\begin{aligned} b(\theta) = b_p(\theta) &= \frac{1}{2-p}[(1-p)\theta]^{\frac{2-p}{1-p}}, p \epsilon (-\infty, 0] \cup (1, \infty), p \neq 1,2, \\ &= \exp\theta, p = 1, \\ &= -\ln(-\theta), p = 2. \end{aligned} \tag{2.11}$$

If $X$ is distributed according to (2.7) with $b(\theta) = b_p(\theta)$, then it will be denoted $X \sim Tw_p^*(\theta, \lambda)$.

A useful alternative form of (2.11) in the case $p \neq 1,2$ is

$$b_p(\theta) = \frac{\alpha - 1}{\alpha}\left[\frac{\theta}{\alpha - 1}\right]^{\alpha}, \tag{2.12}$$

where $\alpha = (2 - p)/(1 - p)$.

**Remark 2.1.** It follows from (2.9)-(2.12) that, for $Tw_p^*$ with $p \neq 1$,

$$E[X] = \lambda\left[\frac{\theta}{\alpha - 1}\right]^{\alpha - 1}. \tag{2.13}$$

$$Var[X] = \lambda\left[\frac{\theta}{\alpha - 1}\right]^{\alpha - 2}, \tag{2.14}$$

and so

$$\theta = (\alpha - 1)\frac{E[X]}{Var[X]}, \tag{2.15}$$

$$\lambda = \frac{Var^{\alpha - 1}[X]}{E^{\alpha - 2}[X]}. \tag{2.16}$$

It also follows from (2.13) and (2.14) that

$$\frac{1}{CoV^2[X]} = \frac{E^2[X]}{Var[X]} = \lambda\left[\frac{\theta}{\alpha - 1}\right]^{\alpha}, \tag{2.17}$$

where $CoV[X]$ denotes the coefficient of variation of $X$. ■

**Remark 2.2.** It may be checked by means of (2.11) that (2.13) and (2.14) continue to hold when $p = 1$ provided that the right-hand sides are interpreted as the limiting case as $\alpha \to \infty$. This yields $E[X] = Var[X] = \lambda e^{\theta}$.

Thus, for fixed $\alpha$, i.e. fixed $p$, Tweedie distributions with the same $\theta$ are those with the same mean-to-variance ratio. ■

**Lemma 2.3.** Suppose that $V_i \sim Tw_p^*(\theta, \lambda_i), i = 1,2$. Then

$$kV_i \sim Tw_p^*(\theta/k, \lambda_i k^{\alpha}), i = 1,2 \text{ for constant } k > 0 \text{ and } p \neq 1, \tag{2.18}$$

$$V_1 + V_2 \sim Tw_p^*(\theta, \lambda_1 + \lambda_2). \tag{2.19}$$

**Proof.** By simple manipulation of the cgf (2.8), using (2.11) and (2.12). ■

**Corollary 2.4.** Suppose that $X \sim Tw_p^*(\theta, \lambda)$. Then (2.18) yields

$$\tilde{X} = X/\lambda \sim Tw_p^*(\lambda\theta, \lambda^{1-\alpha}), \tag{2.20}$$

and then (2.13) and (2.14) yield

$$E[\tilde{X}] = \left[\frac{\theta}{\alpha - 1}\right]^{\alpha - 1}. \tag{2.21}$$

$$Var[\tilde{X}] = \lambda^{-1}\left[\frac{\theta}{\alpha - 1}\right]^{\alpha - 2} = \lambda^{-1}\left[E[\tilde{X}]\right]^{\frac{\alpha - 2}{\alpha - 1}} = \lambda^{-1}\left[E[\tilde{X}]\right]^{p}. \tag{2.22}$$

■

Let the distribution of $\tilde{X}$ be denoted $\tilde{X} \sim Tw_p(\theta, \lambda)$ (no star affixed to $Tw_p$).

**Remark 2.5.** The variate $\tilde{X}$ is the **reproductive representation** of the Tweedie variate $X$. Note that it isolates the parameter $\theta$ from $\lambda$ in (2.21), where it is evident that $\theta$ becomes a location parameter. The reproductive representation is therefore useful in the case where $E[\tilde{X}]$ is to be represented as a specific function of further parameters, such as in a **Generalized Linear Model ("GLM")** (McCullagh and Nelder, 1989). Use will be made of this observation in Section 5.2. ■

#### 2.3.2. Multivariate Tweedie family

Avanzi, Taylor, Vu and Wong (2016) consider variates of the form

$$X_{ij}^{(n)} = \frac{\theta}{\theta_{ij}^{(n)}} W_{ij} + Z_{ij}^{(n)}, \tag{2.23}$$

where $W_{ij} \sim Tw_p^*(\theta, \lambda), Z_{ij}^{(n)} \sim Tw_p^*\left(\theta_{ij}^{(n)}, \lambda_{ij}^{(n)}\right)$ with $p \neq 0$, i.e. $W_{ij}, Z_{ij}^{(n)}$ not normal. The case of normality is simpler and is dealt with in Section 2.3.3.

Then

$$X_{ij}^{(n)} \sim Tw_p^*\left(\theta, \lambda + \lambda_{ij}^{(n)}\right) \text{ for } p = 1, \tag{2.24}$$

$$X_{ij}^{(n)} \sim Tw_p^*\left(\theta, \lambda \left(\frac{\theta}{\theta_{ij}^{(n)}}\right)^{\alpha} + \lambda_{ij}^{(n)}\right) \text{ for } p \neq 1. \tag{2.25}$$

These results may be checked against Lemma 2.3. Since the marginals are Tweedie, the multi-dimensional variate $X_{ij}$ is **multivariate Tweedie**. It should be noted that, if the multiplier of $W_{ij}$ is non-zero and other than that shown in (2.23), then $X_{ij}^{(n)}$ is not Tweedie unless $p = 0$ (normal distribution).

#### 2.3.3. Multivariate log normal family

Let $X$ be a random $n$-vector with **multivariate log normal** distribution, i.e. $ln\, X \sim N(\theta, \Sigma)$, where, here and subsequently, the $ln$ function operates component-wise on its vector argument, $\theta$ is a parameter $n$-vector, and $\Sigma$ is an $n \times n$ covariance matrix. Let the components of $X$ be denoted by $X_k, k = 1, \ldots, n$, similarly for $\theta$, and let the elements of $\Sigma$ be denoted by $\Sigma_{k\ell}, k, \ell = 1, \ldots, n$. Then

$$E[X_k] = exp(\theta_k + ½\Sigma_{kk}), \tag{2.26}$$

$$Cov[X_k, X_\ell] = E[X_k]E[X_\ell](exp\, \Sigma_{k\ell} - 1). \tag{2.27}$$

# 3. Common shock models

## 3.1. Additive common shocks

The general common shock framework of Avanzi, Taylor and Wong (2018) is as follows.

Let $\mathcal{P}^{(n)}$ be a partition of $\mathcal{A}^{(n)} \in \mathbb{A}$, i.e. $\mathcal{P}^{(n)} = \left\{\mathcal{P}_1^{(n)}, \dots, \mathcal{P}_P^{(n)}\right\}$ where the $\mathcal{P}_p^{(n)}$are subsets of $\mathcal{A}^{(n)}$ with $\mathcal{P}_p^{(n)} \cap \mathcal{P}_q^{(n)} = \emptyset$ for all $p, q = 1, \dots, P, p \neq q$ and $\cup_{p=1}^{P} \mathcal{P}_p^{(n)} = \mathcal{A}^{(n)}$. Suppose that all partitions are the same in the sense that, for each $p$, the $(i,j)$ positions of the elements of $\mathcal{A}^{(n)}$ included in $\mathcal{P}_p^{(n)}$ are the same for different $n$. Thus $P$ denotes the order of each $\mathcal{P}^{(n)}$.

Now consider the following dependency structure on the elements $X_{ij}^{(n)}$:

$$X_{ij}^{(n)} = \alpha_{ij}^{(n)} W_{\pi(i,j)} + \beta_{ij}^{(n)} W_{\pi(i,j)}^{(n)} + Z_{ij}^{(n)} \tag{3.1}$$

where $\pi(i,j) = p$ such that $X_{ij}^{(n)} \in \mathcal{P}_p^{(n)}$, a unique mapping; $W_{\pi(i,j)}, W_{\pi(i,j)}^{(n)}, Z_{ij}^{(n)}$ are independent stochastic variates, and $\alpha_{ij}^{(n)}, \beta_{ij}^{(n)} \geq 0$ are fixed and known constants. This independence applies to the case of fixed $i, j, n$. The question of more general dependencies will be considered in Section 3.3.

It is evident that $W_p$ is a common shock across all $n$, but affecting only subsets $\mathcal{P}_p^{(n)}$ for fixed $p$; $W_p^{(n)}$ is similarly a common shock across $\mathcal{P}_p^{(n)}$, but now for fixed $n$ and $p$; and $Z_{ij}^{(n)}$ is an idiosyncratic component of $X_{ij}^{(n)}$, specific to $i, j$ and $n$. The shocks contribute additively to the observation $X_{ij}^{(n)}$.

The common shock $W_p^{(n)}$ creates dependency between observations within the subset $\mathcal{P}_p^{(n)}$ of array $\mathcal{A}^{(n)}$. Since the partitions $\mathcal{P}^{(n)}$ are the same across $n$, the common shock $W_p$ creates dependency between observations in the subsets $\mathcal{P}_p^{(n)}$ of the same or different arrays.

Particular selections of partitions $\mathcal{P}^{(n)}$ are of special interest, as set out in Table 3-1.

**Table 3-1 Special cases of common shock**

| **Type of dependence** | **Partition $\mathcal{P}^{(n)}$** |
|---|---|
| Array-wide | $\left\{\mathcal{P}_1^{(n)}\right\}$ with $\mathcal{P}_1^{(n)} = \mathcal{A}^{(n)}$ |
| Cell-wise | $\left\{\mathcal{P}_1^{(n)}, \mathcal{P}_2^{(n)}, \dots\right\}$ with $\mathcal{P}_1^{(n)} = \left\{X_{11}^{(n)}\right\}, \mathcal{P}_2^{(n)} = \left\{X_{12}^{(n)}\right\}, \dots$ |
| Row-wise | $\left\{\mathcal{P}_1^{(n)}, \dots, \mathcal{P}_I^{(n)}\right\}$ with $\mathcal{P}_i^{(n)} = \mathcal{R}_i^{(n)}$ |
| Column-wise | $\left\{\mathcal{P}_1^{(n)}, \dots, \mathcal{P}_J^{(n)}\right\}$ with $\mathcal{P}_j^{(n)} = \mathcal{C}_j^{(n)}$ |
| Diagonal-wise | $\left\{\mathcal{P}_1^{(n)}, \dots, \mathcal{P}_{I+J-1}^{(n)}\right\}$ with $\mathcal{P}_t^{(n)} = \mathcal{D}_t^{(n)}$ |

## 3.2. Balance of shocks to a claim model

Avanzi, Taylor, Vu and Wong (2021) discussed the issue of balance of an additive common shock model with cell-wise dependence. Taylor and Vu (2022) expanded the discussion to more general dependencies of the form (3.1). Their definition of a balance model required that the proportionate contribution of each shock and idiosyncratic component to cell expectation is constant over all cells, i.e. that $\alpha_{ij}^{(n)} E[W_{\pi(i,j)}]/E[Z_{ij}^{(n)}]$ be independent of $i,j$, and similarly $\beta_{ij}^{(n)} E\left[W_{\pi(i,j)}^{(n)}\right]/E[Z_{ij}^{(n)}]$ be independent of $i,j$.

Avanzi, Taylor, Vu and Wong (2021) introduced an adjustment to achieve rough proportionality in the case of cell-wise dependence between claim triangles. The issue of balance was not addressed in relation to other forms of dependence, but it is clear that different (albeit possibly similar) adjustments would be required.

### 3.3. Multiplicative common shocks

Consider the following alternative to common shock model (3.1):

$$ln\, X_{ij}^{(n)} = \alpha_{ij}^{(n)} ln\, U_{\pi(i,j)} + \beta_{ij}^{(n)} ln\, W_{\pi(i,j)}^{(n)} + ln\, Z_{ij}^{(n)}, \tag{3.2}$$

or, equivalently,

$$X_{ij}^{(n)} = \left(U_{\pi(i,j)}\right)^{\alpha_{ij}^{(n)}} \left(W_{\pi(i,j)}^{(n)}\right)^{\beta_{ij}^{(n)}} Z_{ij}^{(n)}, \tag{3.3}$$

where it is now required that $U_{\pi(i,j)}, W_{\pi(i,j)}^{(n)}, Z_{ij}^{(n)} > 0$.

Let $Z^{(n)}, Z$ be the vectors formed by stacking the $Z_{ij}^{(n)}$ in the same way as the $X_{ij}^{(n)}$ were stacked to form $X^{(n)}, X$. The following independence condition will be assumed.

**Independence assumption 3.1.** It will be required that all vectors $U_{\pi(i,j)}, W_{\pi(i,j)}^{(n)}$ and $Z^{(n)}$ are stochastically independent, though independence is **not** required within these vectors. ■

In $X_{k\ell}^{(n)}$ with $(k,\ell) \in \pi(i,j)$, the multiplier of $Z_{k\ell}^{(n)}$ required to account for the common shock $U_{\pi(i,j)}$ is just the factor $\left(U_{\pi(i,j)}\right)^{\alpha_{k\ell}^{(n)}}$. This is constant over $(k,\ell) \in \pi(i,j)$ if $\alpha_{k\ell}^{(n)} = \alpha_{\pi(i,j)}^{(n)}$ for these $(k,\ell)$. In this case, the effect of the common shock $U_{\pi(i,j)}$ is a uniform multiplier over $\pi(i,j)$. Similar remarks apply to common shock $W_{\pi(i,j)}^{(n)}$. The following proposition compares the balance of additive and multiplicative common shock models.

**Proposition 3.2.** Consider the additive common shock model (3.1) and multiplicative common shock model (3.3) for $X_{k\ell}^{(n)}$, with $(k,\ell) \in \pi(i,j)$. Suppose, in each case, that $\alpha_{k\ell}^{(n)} = \alpha_{\pi(i,j)}^{(n)}, \beta_{k\ell}^{(n)} = \beta_{\pi(i,j)}^{(n)}$, so that $\alpha_{k\ell}^{(n)}, \beta_{k\ell}^{(n)}$ are each constant over $(k,\ell) \in \pi(i,j)$. Then

(a) the multiplier of $Z_{k\ell}^{(n)}$ required to account for the common shock $W_{\pi(i,j)}$ in (3.1) is $\alpha_{\pi(i,j)}^{(n)} W_{\pi(i,j)}/Z_{k\ell}^{(n)}$, which varies inversely with $Z_{k\ell}^{(n)}$;

(b) the multiplier of $Z_{k\ell}^{(n)}$ required to account for the common shock $U_{\pi(i,j)}$ in (3.3) is $\left(U_{\pi(i,j)}\right)^{\alpha_{\pi(i,j)}^{(n)}}$, which is constant over $(k,\ell) \in \pi(i,j)$.

A similar comparison can be made between the contributions of $W_{\pi(i,j)}^{(n)}$ to (3.1) and (3.3) respectively. It follows that the multiplicative model (3.3) is more naturally adapted to balance (in the sense discussed in Section 3.2) than the additive model (3.1). ■

# 4. Log normal claim models

## 4.1. Model structure

Consider a multiplicative common shock model of the form (3.2) with

$$ln\ W_{\pi(i,j)}^{(n)} \sim N\left(\eta_{\pi(i,j)}^{(n)}, \left(\tau_{\pi(i,j)}^{(n)}\right)^2\right), \tag{4.1}$$

$$ln\ U_{\pi(i,j)} \sim N\left(\xi_{\pi(i,j)}, \sigma_{\pi(i,j)}^2\right), \tag{4.2}$$

for given $\eta_{\pi(i,j)}^{(n)}, \left(\tau_{\pi(i,j)}^{(n)}\right)^2, \xi_{\pi(i,j)}, \sigma_{\pi(i,j)}^2$.

The idiosyncratic component $ln\ Z_{ij}^{(n)}$ will usually have some further parametric form (see e.g. Avanzi, Taylor, Vu and Wong (2016)). It will be assumed here that this component is linear in its parameters, which is consistent with the many log-linear claim models in the literature (including the chain ladder). Thus it is assumed that

$$ln\ Z^{(n)} \sim N\left(C^{(n)}\zeta^{(n)}, V^{(n)}\right), \tag{4.3}$$

where

$$ln\ Z^{(n)} = \begin{bmatrix} ln\ Z_{i_1 j_1}^{(n)} \\ ln\ Z_{i_2 j_2}^{(n)} \\ \vdots \end{bmatrix}, \tag{4.4}$$

$\zeta^{(n)}$ is the parameter vector of dimension $q$, and $C^{(n)}$ the design matrix, and where the subscripts $i_1 j_1, i_2 j_2, \ldots$ cover the entirety of each array $\mathcal{A}^{(n)}$.

It is evident from (3.2) and (4.1)-(4.4) that $X$ is log normal. It is convenient to express its distribution in matrix terms. To do so, write

$$ln\ U = \begin{bmatrix} ln\ U_{\pi(i_1,j_1)} \\ ln\ U_{\pi(i_2,j_2)} \\ \vdots \end{bmatrix}, E[U] = \xi, \tag{4.5}$$

where the disjoint union $\pi(i_1,j_1) \cup \pi(i_2,j_2) \cup \ldots$ covers the entirety of each array $\mathcal{A}^{(n)}$.

Similarly,

$$ln\ W^{(n)} = \begin{bmatrix} ln\ W_{\pi(i_1,j_1)}^{(n)} \\ ln\ W_{\pi(i_2,j_2)}^{(n)} \\ \vdots \end{bmatrix}, E\left[ln\ W^{(n)}\right] = \eta^{(n)}, ln\ W = \begin{bmatrix} ln\ W^{(1)} \\ ln\ W^{(2)} \\ \vdots \\ ln\ W^{(N)} \end{bmatrix}, E[ln\ W] = \eta, \tag{4.6}$$

and

$$ln\ Z = \begin{bmatrix} ln\ Z^{(1)} \\ ln\ Z^{(2)} \\ \vdots \\ ln\ Z^{(N)} \end{bmatrix}, E[ln\ Z] = C\zeta, \tag{4.7}$$

with

$$\zeta = \begin{bmatrix} \zeta^{(1)} \\ \zeta^{(2)} \\ \vdots \\ \zeta^{(N)} \end{bmatrix}, C = \begin{bmatrix} C^{(1)} & & & 0 \\ & C^{(2)} & & \\ & & \ddots & \\ 0 & & & C^{(N)} \end{bmatrix}. \tag{4.8}$$

Further, let

$$S = Var[\ln U], \tag{4.9}$$

$$T^{(n)} = \mathrm{Var}\left[\ln W^{(n)}\right], T = Var[\ln W] = \begin{bmatrix} T^{(1)} & & & 0 \\ & T^{(2)} & & \\ & & \ddots & \\ 0 & & & T^{(N)} \end{bmatrix}, \tag{4.10}$$

$$V = Var[\ln Z] = \begin{bmatrix} V^{(1)} & & & 0 \\ & V^{(2)} & & \\ & & \ddots & \\ 0 & & & V^{(N)} \end{bmatrix}. \tag{4.11}$$

The model (3.2) may be expressed in the form

$$Y = A \ln U + B \ln W + \ln Z, \tag{4.12}$$

where $A, B$ are matrices

$$A = \begin{bmatrix} A^{(1)} \\ A^{(2)} \\ \vdots \\ A^{(N)} \end{bmatrix}, B = \begin{bmatrix} B^{(1)} & & & 0 \\ & B^{(2)} & & \\ & & \ddots & \\ 0 & & & B^{(N)} \end{bmatrix}, \tag{4.13}$$

with $A^{(n)}, B^{(n)}$ further matrices. $A^{(n)}$ is a matrix all of whose terms are selected from the $\alpha_{ij}^{(n)}$, and $B^{(n)}$ is a matrix all of whose terms are selected from the $\beta_{ij}^{(n)}$.

It follows from (4.5)-(4.12) that

$$Y \sim N(\theta, \Sigma), \tag{4.14}$$

where

$$\theta = E[Y] = [A \quad B \quad C] \begin{bmatrix} \xi \\ \eta \\ \zeta \end{bmatrix} = M\kappa, \text{say}, \tag{4.15}$$

and

$$\Sigma = Var[Y] = L \begin{bmatrix} S & & 0 \\ & T & \\ 0 & & V \end{bmatrix} L^T, \tag{4.16}$$

with $L = [A \quad B \quad I]$, and (4.16) may then be written as

$$\Sigma = L\Gamma L^T \text{ where } \Gamma = \begin{bmatrix} S & & 0 \\ & T & \\ 0 & & V \end{bmatrix}. \quad (4.17)$$

**Remark 4.1.** The following observations on the model structure may be made.

(a) The vector $Y$ is of dimension $N|\mathcal{A}|$.
(b) The vectors $\xi, \eta, \zeta$ are of dimensions $P, NP$ and $Nq$ respectively.
(c) Dimensions of the matrices involved are as set out in Table 4-1 below.
(d) By assumption, the matrices $T, V$ are block diagonal, but the blocks are, in general, not diagonal. ■

**Table 4-1 Dimensions of matrices**

| Matrix | Dimensions |
|---|---|
| $A$ | $N\|\mathcal{A}\| \times P$ |
| $B$ | $N\|\mathcal{A}\| \times NP$ |
| $C$ | $N\|\mathcal{A}\| \times Nq$ |
| $L$ | $N\|\mathcal{A}\| \times (P + NP + N\|\mathcal{A}\|)$ |
| $M$ | $N\|\mathcal{A}\| \times (P + NP + Nq)$ |
| $\Gamma$ | $(P + NP + N\|\mathcal{A}\|) \times (P + NP + N\|\mathcal{A}\|)$ |
| $\Sigma$ | $N\|\mathcal{A}\| \times N\|\mathcal{A}\|$ |

## 4.2. Parameter estimation

Section 4.1 leads immediately to the following observation.

**Remark 4.2.** In view of (4.14), the model reduces to a **general linear model**, with parameters estimated by **generalized least squares**. ■

There are two cases to be considered according to whether the covariance matrix $\Sigma$ is known or unknown.

### 4.2.1. Covariance matrix $\Sigma$ known

The maximum likelihood **("ML")** estimate of the parameter vector $\kappa$ is

$$\hat{\kappa} = (M^T\Sigma^{-1}M)^{-1}M^T\Sigma^{-1}Y. \quad (4.18)$$

This estimate is unbiased and has covariance matrix

$$Var[\hat{\kappa}] = (M^T\Sigma^{-1}M)^{-1}. \quad (4.19)$$

The **fitted vector** for $Y$, is $\hat{Y} = M\hat{\kappa}$, which is unbiased and has covariance matrix

$$Var[\hat{Y}] = M(M^T\Sigma^{-1}M)^{-1}M^T = \Omega, \text{ say}. \quad (4.20)$$

Since $X = exp\, Y$, an unbiased fitted value for $X_k$, the $k$-th component of $X$, is, by (2.26),

$$\hat{X}_k = exp\left(\hat{Y}_k + ½\Omega_{kk}\right), \tag{4.21}$$

and, by (2.27),

$$Cov\left[\hat{X}_k, \hat{X}_\ell\right] = E\left[\hat{X}_k\right]E\left[\hat{X}_\ell\right](exp\ \Omega_{k\ell} - 1). \tag{4.22}$$

**Remark 4.3.** The relations (4.18)-(4.20) assume that $\Sigma$ is invertible. It is well known that this matrix is not of full rank for certain reserving models, e.g. the chain ladder. When this occurs, it usually results from redundancy in the parameter vector $\theta$. If one formulates the problem as in (4.18)-(4.20), then regression software is likely to return one or more aliased covariates. If one is calculating these results directly from (4.18)-(4.20), then it will be necessary to eliminate redundant parameters from $\theta$, and make corresponding deletions of columns of the matrix $L$. Examples occur in Section7. ■

**Remark 4.4.** The interpretation of the $\hat{X}_k$ in terms of the $\hat{X}_{ij}^{(n)}$, the fitted values corresponding to the $X_{ij}^{(n)}$, requires the unstacking of the vector $\hat{X}$. This is the reverse of the stacking process set out in Section 2.1. First $\hat{X}$ is unstacked into component vectors $\hat{X}^{(n)}$ using the hatted version of (2.1), and then each $\hat{X}^{(n)}$ consists of the $\hat{X}_{ij}^{(n)}$ arranged in the order selected in Section 2.1. ■

#### 4.2.2. Covariance matrix Σ unknown

The case in which $\Sigma$ is unknown may also occur. In this case, some knowledge of the parametric structure of $\Sigma$ will be necessary, as the available data will be insufficient for the estimation of a completely unknown $\Sigma$.

By (4.17), the unknown component of $\Sigma$ is $\Gamma$ in which, by the assumptions of Section 3.3, sub-matrices $T, V$ are block diagonal. Note that the action of the matrix $L$ in (4.17) will scramble the blocks, and so there may be no diagonality in $\Sigma$.

The blocks $V^{(n)}$ in $V$ are unrestricted in Section 3.3, but require restriction here. The simplest assumption would be that each $V^{(n)}$, and therefore $V$, is diagonal. In this case the entire matrix $\Gamma$ could be diagonal but, for the reasons noted in the preceding paragraph, $\Sigma$ might not be.

The general case of structure of $\Gamma$ that enables dimension reduction is $\Gamma = \Gamma(\omega)$, where $\omega$ is a parameter vector. A couple of examples follow.

**Example 4.5.** In the example already given, in which $\Gamma$ is diagonal, $\omega$ is the vector of variances $\sigma_{\pi(i,j)}^2, \left(\tau_{\pi(i,j)}^{(n)}\right)^2, \left(v_{ij,ij}^{(n)}\right)^2$, where $\left(v_{ij,ij}^{(n)}\right)^2$ is the $(ij, ij)$-element of $V^{(n)}$. These restrictions might be tightened further with assumptions that $\sigma_{\pi(i,j)}^2 = \sigma^2, \left(\tau_{\pi(i,j)}^{(n)}\right)^2 = \tau^2, \left(v_{ij,ij}^{(n)}\right)^2 = v^2$, in which case $\Gamma$ is block diagonal with scalar matrices as the blocks:

$$\Gamma = \begin{bmatrix} \sigma^2 I_P & 0 & 0 \\ 0 & \tau^2 I_{NP} & 0 \\ 0 & 0 & v^2 I_{N|\mathcal{A}|} \end{bmatrix}. \tag{4.23}$$

Note that, in the log normal setting, by (4.22), the assumption of a constant dispersion parameter (such as $\sigma^2$) is equivalent to an assumption of constant coefficient of variation over the raw observations. ■

**Example 4.6.** Generalize Example 4.5 in two ways. First, allow the variance parameters to vary with $n$; second, allow the correlation matrices associated with $S, T^{(n)}$ to be non-diagonal but known. That is, assume

$$\Gamma = \begin{bmatrix} \sigma^2 R_U & & & & & & \\ & \left(\tau^{(1)}\right)^2 R_W & & & & & \\ & & \ddots & & & & \\ & & & \left(\tau^{(N)}\right)^2 R_W & & & \\ & & & & \left(v^{(1)}\right)^2 I_{|\mathcal{A}|} & & \\ & & & & & \ddots & \\ & & & & & & \left(v^{(N)}\right)^2 I_{|\mathcal{A}|} \end{bmatrix}. \tag{4.24}$$

where $R_U, R_W$ are the known correlation matrices. ∎

**Example 4.7.** The correlation matrices $R_U, R_W$ in (4.24) are as yet unspecified. A specific example would arise in the case where past data contain a calendar period effect that follows an AR(1) process parallel to the future process (4.51). Thus, the $\pi(i,j)$ represent calendar periods, so that the effect can be present in both the $ln\ U_{\pi(i,j)}$ and the $ln\ W_{\pi(i,j)}^{(n)}$. To be specific, suppose that the respective AR(1) variates are $\gamma_t, \gamma_t^{(n)}$ and all have the same auto-regression coefficient $\rho$ so that, for $t = 1,2, \dots, t_{max}$,

$$\gamma_t = \bar{\gamma} + \rho(\gamma_t - \bar{\gamma}) + \varepsilon_t \tag{4.25}$$

$$\gamma_t^{(n)} = \bar{\gamma}^{(n)} + \rho\left(\gamma_{t-1}^{(n)} - \bar{\gamma}^{(n)}\right) + \varepsilon_t^{(n)} \tag{4.26}$$

Suppose that the AR(1) process is **not** present in the $ln\ Z_{ij}^{(n)}$ and that $V$ continues to be diagonal.

The correlation matrices associated with the $\gamma_t, \gamma_t^{(n)}$ take the form

$$R = \begin{bmatrix} 1 & \rho & \rho^2 & \cdots \\ \rho & 1 & \rho & \cdots \\ \rho^2 & \rho & 1 & \cdots \\ \vdots & \vdots & \vdots & \ddots \end{bmatrix}. \tag{4.27}$$

∎

**Example 4.8.** There is a version of Example 4.6 that is highly simplified but also highly relevant. This is the case in which $R_U = R_W = R$ and $A^{(n)} = B^{(n)} = A^{(0)}$ for $n = 1, \dots, N$, where $A^{(0)}$ is given. This means that the common shock terms $U, W^{(n)}$ are all associated with the same design matrix. In other words, they are all the same, in algebraic structure, in their effects on the $X_{ij}^{(n)}$. The parameters associated with them, components of $\xi$ and $\eta$ can differ, of course.

These conditions greatly simplify the covariance matrices $\Gamma, \Sigma$. By (4.24),

$$\Gamma = \begin{bmatrix} \sigma^2 R & 0 & 0 \\ 0 & \Phi \otimes R & 0 \\ 0 & 0 & \Psi \otimes I_{|\mathcal{A}|} \end{bmatrix}, \tag{4.28}$$

where $\Phi = diag\left(\left(\tau^{(1)}\right)^2, \dots, \left(\tau^{(N)}\right)^2\right), \Psi = diag\left(\left(v^{(1)}\right)^2, \dots, \left(v^{(N)}\right)^2\right)$.

Moreover,

$$L = \left[1_N \otimes A^{(0)}, I_N \otimes A^{(0)}, I_{N|\mathcal{A}|}\right], \qquad (4.29)$$

where, for brevity, $\left|\mathcal{A}^{(n)}\right|$ has been written as just $|\mathcal{A}|$.

By (4.17), (4.28) and (4.29),

$$\Sigma = (\sigma^2 1_N 1_N^T + \Phi) \otimes \left(A^{(0)} R \left(A^{(0)}\right)^T\right) + \Psi \otimes I_{|\mathcal{A}|}, \qquad (4.30)$$

where use has been made of (2.3) to (2.5). ■

Now return to the more general case of Example 4.6. The parameters defining $\Sigma$ are $\sigma^2, \left(\tau^{(1)}\right)^2, \dots, \left(\tau^{(N)}\right)^2, \left(v^{(1)}\right)^2, \dots, \left(v^{(N)}\right)^2$. Let $\omega$ denote the vector of these parameters. To summarize the present situation for that general covariance structure, one is concerned with the following model, from (4.14)-(4.17),

$$Y \sim N\big(M\kappa, \Sigma(\omega)\big), \qquad (4.31)$$

$$\Sigma(\omega) = L\Gamma(\omega)L^T, \qquad (4.32)$$

with the parameter vectors $\kappa, \omega$ requiring estimation.

The log-likelihood of $Y$ is

$$\ell(Y = y) = -½N|\mathcal{A}| \ln 2\pi - ½\ln \det \Sigma - ½(y - M\kappa)^T \Sigma^{-1} (y - M\kappa) \qquad (4.33)$$

where $\Sigma(\omega)$ has been abbreviated to just $\Sigma$.

Then the ML equations are

$$\frac{\partial \ell}{\partial \kappa} = M^T \Sigma^{-1} (y - M\kappa), \qquad (4.34)$$

$$\frac{\partial \ell}{\partial \omega_k} = -\frac{1}{2 \det \Sigma} \frac{\partial \det \Sigma}{\partial \omega_k} - ½(y - M\kappa)^T \frac{\partial \Sigma^{-1}}{\partial \omega_k} (y - M\kappa), \qquad (4.35)$$

where $\omega_k$ is the $k$-th component of the parameter vector $\omega$.

The root of (4.34) is

$$\kappa = (M^T \Sigma^{-1} M)^{-1} M^T \Sigma^{-1} y, \qquad (4.36)$$

from which it follows that the ML estimate $\hat{\kappa}$ of $\kappa$ is just as in (4.18) except that $\Sigma$ must be replaced by $\hat{\Sigma} = \Sigma(\widehat{\omega})$ where $\widehat{\omega}$ is the ML estimate of $\omega$.

Substitution of (4.36) into (4.35) yields

$$\frac{\partial \ell}{\partial \omega_k} = -\frac{1}{2 \det \Sigma} \frac{\partial \det \Sigma}{\partial \omega_k} - ½(y - \hat{y})^T \frac{\partial \Sigma^{-1}}{\partial \omega_k} (y - \hat{y}), \qquad (4.37)$$

where

$$\hat{y} = M\hat{\kappa}. \qquad (4.38)$$

Jacobi's rule may be applied to the member on the right side involving the determinant, converting (4.37) to the following:

$$\frac{\partial \ell}{\partial \omega_k} = -½ \operatorname{tr}\left(\Sigma^{-1}\frac{\partial \Sigma}{\partial \omega_k}\right) - ½(y-\hat{y})^T \frac{\partial \Sigma^{-1}}{\partial \omega_k}(y-\hat{y}), \tag{4.39}$$

with $tr$ denoting the matrix trace.

It is also convenient to express the derivative of $\Sigma^{-1}$ in terms of a derivative of the simpler $\Sigma$. This is achieved by differentiating the identity $\Sigma\Sigma^{-1} = I$ with respect to $\omega_k$ and rearranging slightly to obtain

$$\frac{\partial \ell}{\partial \omega_k} = -½ \operatorname{tr}\left(\Sigma^{-1}\frac{\partial \Sigma}{\partial \omega_k}\right) + ½[\Sigma^{-1}(y-\hat{y})]^T \frac{\partial \Sigma}{\partial \omega_k}[\Sigma^{-1}(y-\hat{y})]. \tag{4.40}$$

ML estimation for the problem (4.31) and (4.32) requires numerical solution for the root $\omega_k$ of (4.40). These roots must be found for all $k$, giving $\hat{\omega}$, and hence $\hat{\Sigma}$. The estimate $\hat{\lambda}$ is then found by means of (4.36).

**Remark 4.9.** The discussion immediately above establishes a protocol for ML estimation in the case of a general covariance structure. In more specific cases, greater guidance can be given on the location of the root of (4.37). Appendix A provides this detail for the specific model described in Example 4.8. ■

## 4.3. Forecast

The model of Section 4.1 can also be used to **forecast** future values of the $X_{ij}^{(n)} \in \mathcal{A}^{(n)*}$, as defined in Section 2.1. The forecasts will be denoted $\hat{X}_{ij}^{(n)}$, and can be stacked into vectors $\hat{X}^{(n)*}$ and $\hat{X}^*$ in the same way as the $X_{ij}^{(n)}$ were stacked to form $X^{(n)}$ and $X$. The upper stars distinguish the vector of predictions from the vector of fitted values.

Section 4.2.1 gave the fitted vector for $Y$ as $\hat{Y} = M\hat{\kappa}$. The corresponding forecast into $\mathcal{A}^{(n)*}, n = 1, \dots, N$ takes the form $\hat{Y}^* = M^*\hat{\kappa}$, for some forecasting matrix $M^*$. Now adopt the notation $\hat{Y}^+$

$$\hat{Y}^+ = \begin{bmatrix} \hat{Y} \\ \hat{Y}^* \end{bmatrix} = \begin{bmatrix} M \\ M^* \end{bmatrix}\hat{\kappa} = M^+\hat{\kappa}. \tag{4.41}$$

In parallel with (4.20), before the addition of process error to forecasts,

$$Var[\hat{Y}^+] = M^+(M^T\Sigma^{-1}M)^{-1}M^{+T}. \tag{4.42}$$

Process error takes the form set out in (4.17) and, when this is included,

$$Var[\hat{Y}^*] = M^*(M^T\Sigma^{-1}M)^{-1}M^{*T} + \Sigma^* = \Omega^*, \text{ say.} \tag{4.43}$$

where $\Sigma^*$ is as in (4.17) but applicable to $Y^*$ rather than $Y$.

In parallel with (4.21) and (4.22),

$$\hat{X}_k^* = exp\left(\hat{Y}_k^* + ½\Omega_{kk}^*\right), \tag{4.44}$$

and

$$Cov\left[\hat{X}_k^*, \hat{X}_\ell^*\right] = E\left[\hat{X}_k^*\right]E\left[\hat{X}_\ell^*\right](exp\Omega_{k\ell}^* - 1). \tag{4.45}$$

Let $\Xi^*$ denote the matrix of these covariances, and let $\Xi^{(n)*}$ denote the $n$-th diagonal block of $\Xi^*$, relating to forecasts in respect of the $n$-th data array.

If a loss reserve is required for claim array $n$, it is constructed as

$$\hat{R}^{(n)} = 1^T\hat{X}^{(n)*}, \tag{4.46}$$

where $\hat{X}^{(n)*}$ may be retrieved from $\hat{X}^+$; and the total reserve across all arrays is

$$\hat{R} = \sum_{n=1}^{N} \hat{R}^{(n)} = 1^T\hat{X}^* . \tag{4.47}$$

It then follows that

$$Var\left[\hat{R}^{(n)}\right] = 1^T\Xi^{(n)*}1, \tag{4.48}$$

$$Var\left[\hat{R}\right] = 1^T\Xi^*1, \tag{4.49}$$

*Parameters specific to future claim experience*

Some forecasts may require parameters specific to future claim experience. The most obvious example is that incorporating a **calendar period effect**. Suppose, for example that $Z_{ij}^{(n)}$ is modelled in such a way that

$$E\left[Z_{ij}^{(n)}\right] = a_i^{(n)} + f^{(n)}(j;b) + \gamma_t^{(n)}, \tag{4.50}$$

where the $\gamma_t^{(n)}$ describe the calendar period effect, the $a_i^{(n)}$ are accident period parameters, and $f^{(n)}$ is a function, dependent on parameter vector $b$, that describes the development period effect.

Application of ML estimation to the data will provide estimates $\hat{\gamma}_t^{(n)}$ of the $\gamma_t^{(n)}$ for $t \le t_{max}$. However, the future observations requiring forecast all depend on $\gamma_t^{(n)}, t > t_{max}$. These cannot be estimated from the available data, but must be inserted into the forecast "by hand".

For example, future values of $\gamma_t^{(n)}$ might be assumed generated by an AR(1) process:

$$\gamma_t^{(n)} = \bar{\gamma}^{(n)} + \rho^{(n)}\left(\gamma_{t-1}^{(n)} - \bar{\gamma}^{(n)}\right) + \varepsilon_t^{(n)}, t = t_{max} + 1, t_{max}, +2, ..., \tag{4.51}$$

Where $\varepsilon_t^{(n)}$ is a centred stochastic error, $\bar{\gamma}^{(n)}$ and $\rho^{(n)}$ are constants specific to $n$, and the series is initiated by $\gamma_{t_{max}}^{(n)} = \hat{\gamma}_{t_{max}}^{(n)}$.

# 5. Log Tweedie models

## 5.1. Statistical properties of Tweedie common shock models

It is possible to generalize the log normal model of Section 4 to a **log Tweedie common shock model**, i.e. a model in which logged observations are Tweedie distributed. The model continues to follow the construction (3.2), but now

$$ln\, U_{\pi(i,j)} \sim Tw_p^*\left(\theta_{\pi(i,j)}, \lambda_{\pi(i,j)}\right), \tag{5.1}$$

$$ln\, W_{\pi(i,j)}^{(n)} \sim Tw_p^*\left(\theta_{\pi(i,j)}^{(n)}, \lambda_{\pi(i,j)}^{(n)}\right), \tag{5.2}$$

$$ln\, Z_{ij}^{(n)} \sim Tw_p^*\left(\theta_{ij}^{(n)}, \lambda_{ij}^{(n)}\right), \tag{5.3}$$

for given parameters $\theta_{\pi(i,j)}, \lambda_{\pi(i,j)}, \theta_{\pi(i,j)}^{(n)}, \lambda_{\pi(i,j)}^{(n)}, \theta_{ij}^{(n)}, \lambda_{ij}^{(n)}$, and with the following independence assumption.

**Independence assumption 5.1.** It is assumed that the variates $ln\, U_{\pi(i_1,j_1)}, ln\, W_{\pi(i_2,j_2)}^{(n)}, ln\, Z_{i_3,j_3}^{(m)}$ are independent across all $n, m, i_1, j_1, i_2, j_2, i_3, j_3$. Note that this is a stronger restriction than Independence assumption 3.1. ■

As noted in Section 2.3.2 (see particularly (2.23)), the coefficients associated with the random variables (5.1) to (5.3) are not free in (3.2). The admissible forms are given by Taylor and Vu (2022) as

$$\alpha_{ij}^{(n)} = \theta_{\pi(i,j)} / \theta_{ij}^{(n)}, \tag{5.4}$$

$$\beta_{ij}^{(n)} = \theta_{\pi(i,j)}^{(n)} / \theta_{ij}^{(n)}. \tag{5.5}$$

In view of these restrictions, the log normal model of Section 4 cannot be extracted as special case of the log Tweedie developed in the present section.

It is also shown by Taylor and Vu (2022) that

$$ln\, X_{ij}^{(n)} \sim Tw_p^*\left(\theta_{ij}^{(n)}, \phi_{ij}^{(n)}\right). \tag{5.6}$$

with

$$\phi_{ij}^{(n)} = \left(\frac{\theta_{\pi(i,j)}}{\theta_{ij}^{(n)}}\right)^{\alpha} \lambda_{\pi(i,j)} + \left(\frac{\theta_{\pi(i,j)}^{(n)}}{\theta_{ij}^{(n)}}\right)^{\alpha} \lambda_{\pi(i,j)}^{(n)} + \lambda_{ij}^{(n)}. \tag{5.7}$$

In fact, the three components of the parameter $\phi_{ij}^{(n)}$ are the corresponding parameters of the three components of $Y_{ij}^{(n)}$. The formal statement of this is as follows.

**Proposition 5.2.** In the model (3.2), subject to (5.1)-(5.5) and independence assumption 5.1, the three components of (3.2) are independent and are distributed as follows:

$$\alpha_{ij}^{(n)} ln\, U_{\pi(i,j)} \sim Tw_p^*\left(\theta_{ij}^{(n)}, \left(\frac{\theta_{\pi(i,j)}}{\theta_{ij}^{(n)}}\right)^{\alpha} \lambda_{\pi(i,j)}\right), \tag{5.8}$$

$$\beta_{ij}^{(n)} ln\, W_{\pi(i,j)}^{(n)} \sim Tw_p^*\left(\theta_{ij}^{(n)}, \left(\frac{\theta_{\pi(i,j)}^{(n)}}{\theta_{ij}^{(n)}}\right)^{\alpha} \lambda_{\pi(i,j)}^{(n)}\right), \tag{5.9}$$

$$ln\, Z_{ij}^{(n)} \sim Tw_p^*\left(\theta_{ij}^{(n)}, \lambda_{ij}^{(n)}\right). \tag{5.10}$$

**Proof.** By Lemma 2.3. ■

The following notation will be useful:

$$u_{ij}^{(n)} = \left\{\frac{\lambda_{\pi(i,j)}}{\lambda_{ij}^{(n)}}\left(\frac{\theta_{\pi(i,j)}}{\theta_{ij}^{(n)}}\right)^{\alpha}\right\}^{½} = \frac{CoV\left[ln\, Z_{ij}^{(n)}\right]}{CoV\left[ln\, U_{\pi(i,j)}\right]},$$

$$w_{ij}^{(n)} = \left\{\frac{\lambda_{\pi(i,j)}^{(n)}}{\lambda_{ij}^{(n)}}\left(\frac{\theta_{\pi(i,j)}^{(n)}}{\theta_{ij}^{(n)}}\right)^{\alpha}\right\}^{½} = \frac{CoV\left[ln\, Z_{ij}^{(n)}\right]}{CoV\left[ln\, W_{\pi(i,j)}^{(n)}\right]},$$

where the right-most equalities follow from (2.17).

The parameter $\phi_{ij}^{(n)}$ in (5.7) may be re-expressed as

$$\phi_{ij}^{(n)} = \lambda_{ij}^{(n)}\left\{\frac{\lambda_{\pi(i,j)}\left(\theta_{\pi(i,j)}\right)^{\alpha}}{\lambda_{ij}^{(n)}\left(\theta_{ij}^{(n)}\right)^{\alpha}} + \frac{\lambda_{\pi(i,j)}^{(n)}\left(\theta_{\pi(i,j)}^{(n)}\right)^{\alpha}}{\lambda_{ij}^{(n)}\left(\theta_{ij}^{(n)}\right)^{\alpha}} + 1\right\} = \lambda_{ij}^{(n)}\psi_{ij}^{(n)}\,. \tag{5.11}$$

where

$$\psi_{ij}^{(n)} = 1 + \left(u_{ij}^{(n)}\right)^2 + \left(w_{ij}^{(n)}\right)^2. \tag{5.12}$$

**Proposition 5.3.** Combining (5.11) with (5.6) and (5.7) yields

$$Y_{ij}^{(n)} \sim Tw_p^*\left(\theta_{ij}^{(n)}, \lambda_{ij}^{(n)}\psi_{ij}^{(n)}\right). \tag{5.13}$$ ■

Proposition 5.2 may be applied to the evaluation of expectations, variances and covariances of the $Y_{ij}^{(n)}$.

It follows from (2.13) and (5.3) that

$$E\left[ln\, Z_{ij}^{(n)}\right] = \lambda_{ij}^{(n)}\left(\frac{\theta_{ij}^{(n)}}{\alpha - 1}\right)^{\alpha-1}, \tag{5.14}$$

and similarly, from (2.13) and (5.13), that

$$E\left[Y_{ij}^{(n)}\right] = \lambda_{ij}^{(n)}\psi_{ij}^{(n)}\left(\frac{\theta_{ij}^{(n)}}{\alpha - 1}\right)^{\alpha-1} = \psi_{ij}^{(n)}E\left[ln\, Z_{ij}^{(n)}\right], \tag{5.15}$$

By (2.14), (5.8) and (5.10),

$$Var\left[ln\, Z_{ij}^{(n)}\right] = \lambda_{ij}^{(n)}\left(\frac{\theta_{ij}^{(n)}}{\alpha - 1}\right)^{\alpha-2}, \tag{5.16}$$

$$\begin{aligned} Var\left[\alpha_{ij}^{(n)}ln\, U_{\pi(i,j)}\right] &= \lambda_{\pi(i,j)}\left(\frac{\theta_{\pi(i,j)}}{\theta_{ij}^{(n)}}\right)^{\alpha}\left(\frac{\theta_{ij}^{(n)}}{\alpha - 1}\right)^{\alpha-2} \\ &= \frac{\lambda_{\pi(i,j)}}{\lambda_{ij}^{(n)}}\left(\frac{\theta_{\pi(i,j)}}{\theta_{ij}^{(n)}}\right)^{\alpha} Var\left[ln\, Z_{ij}^{(n)}\right] = \left(u_{ij}^{(n)}\right)^2 Var\left[ln\, Z_{ij}^{(n)}\right]. \end{aligned} \tag{5.17}$$

Similarly,

$$Var\left[\beta_{ij}^{(n)} ln\, W_{\pi(i,j)}^{(n)}\right] = \left(w_{ij}^{(n)}\right)^2 Var\left[ln\, Z_{ij}^{(n)}\right]. \tag{5.18}$$

For covariances, consider the quantity

$$\begin{aligned}\left\{\alpha_{ij}^{(n)}\alpha_{ij}^{(m)} Var\left[ln\, U_{\pi(i,j)}\right]\right\}^2 &= Var\left[\alpha_{ij}^{(n)} ln\, U_{\pi(i,j)}\right] Var\left[\alpha_{ij}^{(m)} ln\, U_{\pi(i,j)}\right] \\ &= \left(u_{ij}^{(n)}u_{ij}^{(m)}\right)^2 Var\left[ln\, Z_{ij}^{(n)}\right] Var\left[ln\, Z_{ij}^{(m)}\right] \quad \text{by (5.17)},\end{aligned} \tag{5.19}$$

and so

$$\alpha_{ij}^{(n)}\alpha_{ij}^{(m)} Var\left[ln\, U_{\pi(i,j)}\right] = u_{ij}^{(n)}u_{ij}^{(m)} \left\{Var\left[ln\, Z_{ij}^{(n)}\right] Var\left[ln\, Z_{ij}^{(m)}\right]\right\}^{1/2}. \tag{5.20}$$

Similarly,

$$\beta_{ij}^{(n)}\beta_{ij}^{(m)} Var\left[ln\, W_{\pi(i,j)}^{(n)}\right] = w_{ij}^{(n)}w_{ij}^{(m)} \left\{Var\left[ln\, Z_{ij}^{(n)}\right] Var\left[ln\, Z_{ij}^{(m)}\right]\right\}^{1/2}. \tag{5.21}$$

To calculate the covariance of $Y_{ij}^{(n)}$ and $Y_{k\ell}^{(m)}$, recall that all $U, W, Z$ variates are independent, whereupon results (5.15), (5.17), (5.18), (5.20) and (5.21) yield the following proposition.

**Proposition 5.4.** Let $\delta$ denote the Kronecker delta that takes unit value when its two subscripts are identical, and value zero otherwise. Then

$$E\left[Y_{ij}^{(n)}\right] = \psi_{ij}^{(n)} E\left[ln\, Z_{ij}^{(n)}\right], \tag{5.22}$$

$$Var\left[Y_{ij}^{(n)}\right] = \psi_{ij}^{(n)} Var\left[ln\, Z_{ij}^{(n)}\right], \tag{5.23}$$

$$\begin{aligned}Cov&\left[Y_{ij}^{(n)}, Y_{k\ell}^{(m)}\right] \\ &= \left(\delta_{\pi(i,j),\pi(k,\ell)} u_{ij}^{(n)}u_{ij}^{(m)} + \delta_{n,m}\delta_{\pi(i,j),\pi(k,\ell)} w_{ij}^{(n)}w_{ij}^{(m)} \right. \\ &\quad \left. + \delta_{n,m}\delta_{(i,j),(k,\ell)}\right)\left\{Var\left[ln\, Z_{ij}^{(n)}\right] Var\left[ln\, Z_{ij}^{(m)}\right]\right\}^{1/2},\end{aligned} \tag{5.24}$$

$$\begin{aligned}Corr&\left[Y_{ij}^{(n)}, Y_{k\ell}^{(m)}\right] \\ &= \frac{\delta_{\pi(i,j),\pi(k,\ell)} u_{ij}^{(n)}u_{ij}^{(m)} + \delta_{n,m}\delta_{\pi(i,j),\pi(k,\ell)} w_{ij}^{(n)}w_{ij}^{(m)} + \delta_{n,m}\delta_{(i,j),(k,\ell)}}{\left\{\psi_{ij}^{(n)}\psi_{ij}^{(m)}\right\}^{1/2}}.\end{aligned} \tag{5.25}$$

■

**Remark 5.5.** These covariances and correlations will be helpful in implementing the model (3.2) and (5.1)-(5.5). It is convenient that the correlations require no quantities other than the ratios of CoVs, $u_{ij}^{(n)}$ and $w_{ij}^{(n)}$, as the user may be able to guess reasonable values for them. ■

**Remark 5.6.** A particularly simple, but not wholly unrealistic, case arises when $CoV\left[ln\, U_{\pi(i,j)}\right] = CoV\left[ln\, W_{\pi(i,j)}^{(n)}\right] = CoV\left[ln\, Z_{ij}^{(n)}\right]$. Then $u_{ij}^{(n)} = w_{ij}^{(n)} = 1$, and (5.25) reduces to

$$Corr\left[Y_{ij}^{(n)}, Y_{k\ell}^{(m)}\right] = \tfrac{1}{3}\left(\delta_{\pi(i,j),\pi(k,\ell)} + \delta_{n,m}\delta_{\pi(i,j),\pi(k,\ell)} + \delta_{n,m}\delta_{(i,j),(k,\ell)}\right). \tag{5.26}$$

■

**Example 5.7.** Consider the case of row-wise dependence described in Table 3-1, where $\delta_{\pi(i,j),\pi(k,\ell)} = \delta_{i,k}$, whence (5.25) reduces to

$$Corr\left[Y_{ij}^{(n)}, Y_{k\ell}^{(m)}\right] = \frac{\delta_{i,k} u_{ij}^{(n)} u_{ij}^{(m)} + \delta_{n,m}\delta_{i,k} w_{ij}^{(n)} w_{ij}^{(m)} + \delta_{n,m}\delta_{(i,j),(k,\ell)}}{\left\{\psi_{ij}^{(n)}\psi_{ij}^{(m)}\right\}^{½}}. \quad (5.27) ■$$

## 5.2. Model structure and parameter estimation

The model construction of Section 4.1 can be translated to the context of Section 5.1. The same notational conventions are used for stacking vectors and matrices. Equations (4.5)-(4.7) and (4.12) continue to hold.

The assumption made by Avanzi, Taylor, Vu and Wong (2016) was a special case of (4.7). The choice of matrices $A$ and $B$ in (4.12) is no longer free, but restricted by (5.4) and (5.5). The covariance matrices (4.9)-(4.11) continue to be free except that they are now, by assumption, diagonal.

Define the scaled version of $Y_{ij}^{(n)}$,

$$\breve{Y}_{ij}^{(n)} = Y_{ij}^{(n)}/\psi_{ij}^{(n)}, \quad (5.28)$$

whence

$$E\left[\breve{Y}_{ij}^{(n)}\right] = E\left[\ln Z_{ij}^{(n)}\right] = \left[c_{ij}^{(n)}\right]^T \zeta^{(n)}, \quad (5.29)$$

by (5.22) and then (4.7), where $\left[c_{ij}^{(n)}\right]^T$ is the $(i,j)$-th row of design matrix $C^{(n)}$.

More generally, by Lemma 2.3 applied to (5.28), and taking account of (5.13),

$$\breve{Y}_{ij}^{(n)} \sim Tw_p^*\left(\theta_{ij}^{(n)}\psi_{ij}^{(n)}, \lambda_{ij}^{(n)}\left(\psi_{ij}^{(n)}\right)^{1-\alpha}\right), \quad (5.30)$$

whence (2.14) yields

$$Var\left[\breve{Y}_{ij}^{(n)}\right] = \lambda_{ij}^{(n)}\left(\psi_{ij}^{(n)}\right)^{1-\alpha}\left(\frac{\theta_{ij}^{(n)}\psi_{ij}^{(n)}}{\alpha-1}\right)^{\alpha-2} = \frac{\left(\lambda_{ij}^{(n)}\right)^{1-p}}{\psi_{ij}^{(n)}}\left(E\left[\breve{Y}_{ij}^{(n)}\right]\right)^p, \quad (5.31)$$

after a small amount of manipulation and recognition of the relation between $\alpha$ and $p$.

The following proposition summarizes the situation.

**Proposition 5.7.** By (5.29)-(5.31), the observations $\breve{Y}_{ij}^{(n)}$ would, but for the correlations (5.25), be represented by a GLM, with Tweedie parameter $p$, mean $\left[c_{ij}^{(n)}\right]^T \zeta^{(n)}$, identity link function, unit scale parameter and weights $\left(\lambda_{ij}^{(n)}\right)^{p-1}\psi_{ij}^{(n)}$. The existence of non-zero correlations (5.25) simply converts the GLM into a system of **Generalized Estimating Equations (GEE)** (Liang and Zeger, 1986), embedding the above GLM structure of each observation. ■

It is noted that an R package (Halekoh and Højsgaard, 2006) is available for the treatment of GEEs.

# 6. Parametric model structures

The GEE established in Proposition 5.7 has the very general structure in which the vector of observation means has the general linear form (5.29). The covariates involved are as yet abstract. The next two sub-sections consider some particular cases. Some of this material is drawn largely from Taylor (2000, Chapter 8) and Taylor and McGuire (2016).

## 6.1. Chain ladder

The most familiar loss reserving model is the **chain ladder**. This model exists in a number of versions (Taylor, 2011), of which some are closely adapted to the models discussed in this paper. These are the so-called **cross-classified** models, which all have the following characteristic:

$$E\left[X_{ij}^{(n)}\right] = exp\left(\chi'^{(n)}_i\right)exp\left(\rho'^{(n)}_j\right), \tag{6.1}$$

for row-effect and column-effect parameters $\chi'^{(n)}_i$ and $\rho'^{(n)}_j$ respectively.

### 6.1.1. Log normal chain ladder

A log normal version of this model is given by Wüthrich and Merz (2008, Section 5.2.3). If the $Y_{ij}^{(n)}$ for a single $n$ are made the subject of this model, then it may be stated as

$$Y_{ij}^{(n)} \sim N\left(\mu_{ij}^{(n)}, \sigma^{2(n)}\right), \tag{6.2}$$

with

$$\mu_{ij}^{(n)} = c^{(n)} + \chi_i^{(n)} + \rho_j^{(n)} \tag{6.3}$$

where $c^{(n)}, \chi_i^{(n)}, \rho_j^{(n)}$ are further parameters.

The log normal common shock model of Section 4.1 can be compared with the above Wüthrich-Merz model. By (4.15),

$$E\left[Y_{ij}^{(n)}\right] = \left[a_{ij}^{(n)}\right]^T \xi + \left[b_{ij}^{(n)}\right]^T \eta^{(n)} + \left[c_{ij}^{(n)}\right]^T \zeta^{(n)}, \tag{6.4}$$

where $\left[a_{ij}^{(n)}\right]^T, \left[b_{ij}^{(n)}\right]^T, \left[c_{ij}^{(n)}\right]^T$ are the $(i,j)$-th rows of design matrices $A^{(n)}, B^{(n)}, C^{(n)}$.

The idiosyncratic parameter vector $\zeta^{(n)}$ can be chosen to consist of row-specific scalars $\chi_i^{(n)}$ and column-specific scalars $\rho_j^{(n)}$. With $C^{(n)}$ consisting of suitably placed unit and zero entries, the final summand of (6.4) takes the form

$$\left[c_{ij}^{(n)}\right]^T \zeta^{(n)} = \chi_i^{(n)} + \rho_j^{(n)}. \tag{6.5}$$

Then

$$Y_{ij}^{(n)} \sim N\left(\mu_{ij}^{(n)}, \sigma_{ij}^{2(n)}\right), \tag{6.6}$$

with

$$\mu_{ij}^{(n)} = \left[a_{ij}^{(n)}\right]^T \xi + \left[b_{ij}^{(n)}\right]^T \eta^{(n)} + \chi_i^{(n)} + \rho_j^{(n)} \tag{6.7}$$

and

$$\sigma_{ij}^{2(n)} = \Omega_{ij,ij}^{(nn)} \tag{6.8}$$

where, in general, $\Omega_{ij,k\ell}^{(nm)} = Cov\left[Y_{ij}^{(n)}, Y_{k\ell}^{(m)}\right]$.

Then comparison of (6.6)-(6.8) with (6.2) and (6.3) reveals that the log normal common shock model is identical to the log normal chain ladder model supplemented by the common shocks.

### 6.1.2. Log Tweedie chain ladder

Reasoning parallel to that of Section 6.1.1 may be applied to produce a log Tweedie common shock model. One commences with (5.22) and (4.7) to obtain

$$E\left[Y_{ij}^{(n)}\right] = \psi_{ij}^{(n)}\left[c_{ij}^{(n)}\right]^T \zeta^{(n)}\ . \tag{6.9}$$

Once again, one chooses the chain ladder form (6.5) to yield

$$E\left[Y_{ij}^{(n)}\right] = \psi_{ij}^{(n)}\left(\chi_i^{(n)} + \rho_j^{(n)}\right). \tag{6.10}$$

Note that, if $\psi_{ij}^{(n)}$ assumes the simplified form $\psi^{(n)}$, then the $\psi$ parameters may be absorbed within the $\chi_i^{(n)}, \rho_j^{(n)}$, and the previous expression reduces to just

$$E\left[Y_{ij}^{(n)}\right] = \chi_i^{(n)} + \rho_j^{(n)}\ . \tag{6.11}$$

Then

$$E\left[X_{ij}^{(n)}\right] = E\left[exp\ Y_{ij}^{(n)}\right] = exp\left(\chi_i^{(n)}\right) exp\left(\rho_j^{(n)}\right) + 2nd\ order\ adjustment, \tag{6.12}$$

when the Taylor expansion of the expectation is taken.

Note that the correspondence with the chain ladder model is a little more direct here than in the log normal case of Section 6.1.1. No common shock adjustment terms are required in (6.11) as, according to (5.22), on which the reasoning just above is based, these are just multiples of the idiosyncratic term. This arises from the action of the restrictions (5.4) and (5.5) on the model form (3.2) in the log Tweedie case.

## 6.2. More general models

The log normal chain ladder form in Section 6.1.1 was defined essentially by (6.5). Other parametric forms may be inserted into this relation, e.g.

$$\left[c_{ij}^{(n)}\right]^T \zeta^{(n)} = c_i^{(n)} + \chi^{(n)} ln\, j - \rho^{(n)} j, \tag{6.13}$$

so that (6.6) holds but now with

$$\mu_{ij}^{(n)} = \left[a_{ij}^{(n)}\right]^T \xi + \left[b_{ij}^{(n)}\right]^T \eta^{(n)} + c_i^{(n)} + \chi^{(n)} ln\, j - \rho^{(n)} j. \tag{6.14}$$

It follows that

$$\begin{aligned} E\left[X_{ij}^{(n)}\right] &= E\left[exp\ Y_{ij}^{(n)}\right] \\ &= \left(exp\ c_i^{(n)}\right) j^{\chi^{(n)}} \exp\left(-\rho^{(n)} j\right) \\ &\times common\ shock\ adjustments \times variance\ adjustment. \end{aligned} \tag{6.15}$$

The term before the adjustment factors on the right can be recognised as a Hoerl curve in $j$, with different amplitudes for different $i$. This parametric form occurs in De Jong and Zehnwirth (1983) and Taylor (2000, Chapter 8).

It is not difficult to produce other parametric forms in place of (6.13). The log Tweedie model of Section 6.1.2 may also be extended to these in place of the log normal.

# 7. Numerical example: cell-wise dependence across triangles

## 7.1. Model structure

Avanzi, Taylor, Vu and Wong (2016) investigated a numerical case in which a cell-wise common shock across two segments was assumed ($N = 2$). A similar case will be analyzed here using the log normal model of Section 4.2.2.

The model is a special case of (4.12) but, since it includes no within-segment common shock, the $W$ term is eliminated, and the model becomes

$$Y = A\, ln\, U + ln\, Z, \tag{7.1}$$

According to Table 3-1, the partition $\mathcal{P}^{(n)}$ of data segment $\mathcal{A}^{(n)}$ consists of all the individual observations in $\mathcal{A}^{(n)}$. Therefore, (4.5) yields

$$ln\, U = \begin{bmatrix} ln\, U_{11} \\ ln\, U_{12} \\ \vdots \end{bmatrix},$$

the components covering all cells of any one $\mathcal{A}^{(n)}$. It follows that $P = |\mathcal{A}|$.

Apply a simplified version of Example 4.8 in which the $ln\, U_{ij}$ are equi-distributed and independent, the middle row and column of blocks are deleted from (4.23), and $A^{(0)} = I_{|\mathcal{A}|}, R = I_{|\mathcal{A}|}, v^{(1)} = \cdots = v^{(N)} = v$, so that in (4.28) $\Psi \otimes I_{|\mathcal{A}|} = v^2 I_{\mathrm{N}} \otimes I_{|\mathcal{A}|} = v^2 I_{N|\mathcal{A}|}$. The omission of the within-array common shocks may be recognized by setting $\Phi = 0$ in (4.30).

Substitution of all of these quantities into (4.30) yields

$$\Sigma = \sigma^2 (1_N 1_N^T) \otimes I_{|\mathcal{A}|} + v^2 I_{N|\mathcal{A}|}. \tag{7.2}$$

The regression design (4.15) now becomes

$$\theta = M\kappa, \tag{7.3}$$

with

$$M = \begin{bmatrix} 1_N \otimes I_{|\mathcal{A}|} & C \end{bmatrix}, \kappa = \begin{bmatrix} \xi 1_{|\mathcal{A}|} \\ \zeta \end{bmatrix}, \tag{7.4}$$

where $\xi$ is now a scalar.

Finally, it will be assumed that the idiosyncratic components $Z_{ij}^{(n)}$ assume the chain ladder structure

$$E\left[Z_{ij}^{(n)}\right] = \chi_i^{(n)} + \rho_j^{(n)}, \tag{7.5}$$

for parameters $\chi_i^{(n)}, \rho_j^{(n)}$.

This structure will be captured by (7.4) if $\zeta, C$ are defined by

$$\zeta^{(n)} = \begin{bmatrix} \chi^{(n)} \\ \rho^{(n)} \end{bmatrix}, C^{(n)} = \begin{bmatrix} 1_{\left|\mathcal{R}_1^{(n)}\right|} e_1^T & E_1 \\ 1_{\left|\mathcal{R}_2^{(n)}\right|} e_2^T & E_2 \\ \vdots & \vdots \end{bmatrix}, \tag{7.6}$$

where $\chi^{(n)}$ is the vector of all $\chi_i^{(n)}$ and $b^{(n)}$ the vector of all $\rho_j^{(n)}$, in each case in ascending order of $i$ or $j$, as the case may be. Here, $C^{(n)}$ is a block matrix, $e_i$ denotes the $i$-th natural coordinate vector of dimension $J$, and, for given $i$, $E_i$ is a selector matrix of dimension $\left|\mathcal{R}_i^{(n)}\right| \times J$ with each row a natural coordinate vector $e_j$ when the observation $X_{ij}^{(n)}$ exists.

## 7.2. Data

Two synthetic claim rectangles, each of dimension 15, were generated in accordance with the model set out in Section 7.1. Selected values of the parameters are as in Table 7-1 *et seq*. Note that, because of (3.2), the parameters are more informatively expressed in the exponentiated form shown. Not also that the two triangles are highly imbalanced, one with quite long tail, and the other quite short.

**Table 7-1 Model parameter values**

| j | $exp\ \rho_j^{(1)}$ | $exp\ \rho_j^{(2)}$ |
|---|---|---|
| 1 | 0.02 | 0.10 |
| 2 | 0.03 | 0.40 |
| 3 | 0.05 | 0.30 |
| 4 | 0.10 | 0.15 |
| 5 | 0.15 | 0.03 |
| 6 | 0.15 | 0.01 |
| 7 | 0.12 | 0.005 |
| 8 | 0.10 | 0.0025 |
| 9 | 0.08 | 0.0010 |
| 10 | 0.07 | 0.0005 |
| 11 | 0.05 | 0.0002 |
| 12 | 0.04 | 0.0002 |
| 13 | 0.025 | 0.0002 |
| 14 | 0.010 | 0.0002 |
| 15 | 0.005 | 0.0002 |

Other parameter values are:

$exp\ \chi_i^{(1)} = 10000 \times e^{0.02(i-1)}, exp\ \chi_i^{(2)} = 30000 \times e^{-0.02(i-1)}, \xi = 0.15, \sigma = 0.1, v = 0.15$.

The simulated rectangles appear in Appendix C.

### 7.3. Parameter estimation and forecast

Substitution of (7.6) into (7.3) and (7.4) provides a somewhat more explicit model statement, as follows:

$$\theta = \xi 1_{N|\mathcal{A}|} + C\zeta, \tag{7.7}$$

where the first vector on the right has been obtained by the use of (2.5). Relation (7.7) illustrates how the common shock expectation $\xi$ affects all cells across $i, j, n$.

At this point, one may usefully note redundancy in model definition (7.7). The typical component of vector $\theta$ may be expressed in the form:

$$u + \chi_i^{(n)} + \rho_j^{(n)} = \left(\chi_i^{(n)} - \chi_1^{(n)}\right) + \left(u + \chi_1^{(n)} + \rho_j^{(n)}\right) = \chi_i^{(n)} + \left(u + \rho_j^{(n)}\right), \tag{7.8}$$

if a constraint $\chi_1^{(n)} = 0$ is applied. In this case, the cell mean on the left side of (7.8) can be expressed as a row effect $\chi_i^{(n)}, i = 2, \dots, 15$ (with $\chi_1^{(n)} = 0$) and a column effect $\left(u + \rho_j^{(n)}\right), j = 1, \dots, 15$, i.e. 29 parameters. Note that the common shock $u$ has been absorbed into the column effect. This means that, in parameter estimation, the common shock cannot be disentangled from the chain ladder structure. However, it will be incorporated in any forecasts, and its effect on the variance-covariance structure can be estimated.

The redundancies identified here are such that:

- The first of the two members on the right of (7.7) can be deleted, and (7.7) replaced by:

  $$\theta = C\zeta, \tag{7.9}$$

- The vector $\zeta^{(n)}$ is reduced so that its components are now $\chi_i^{(n)}, i = 2, \dots, 15, \left(u + \rho_j^{(n)}\right), j = 1, \dots, 15$.
- The first column of $C^{(n)}$ can be deleted since $\chi_1^{(n)}$ no longer enters into the estimation problem.

The estimation problem now becomes generalized least squares regression with design given by (4.14), (7.2) and (7.9). The solution is given in Section 4.2.2 (see (4.34)-(4.36)). This can be adapted to the present situation, using (7.2) to obtain relevant expressions for $\Sigma^{-1}$ and $\partial\Sigma/\partial\omega_k$.

One may check that

$$\Sigma^{-1} = \frac{1}{v^2(2\sigma^2 + v^2)} \begin{bmatrix} (\sigma^2 + v^2)I & -\sigma^2 I \\ -\sigma^2 I & (\sigma^2 + v^2)I \end{bmatrix}, \tag{7.10}$$

$$\frac{\partial\Sigma}{\partial\sigma^2} = \begin{bmatrix} I & I \\ I & I \end{bmatrix}, \frac{\partial\Sigma}{\partial v^2} = \begin{bmatrix} I & 0 \\ 0 & I \end{bmatrix}. \tag{7.11}$$

Substitution of these quantities into (4.40) leads to the following result. Details of the calculation appear in Appendix B.

**Proposition 7.1.** Set $r = \sigma^2/v^2$. Then the ML estimate (4.40) of $v^2$ is

$$\hat{v}^2 = \frac{\left(d^{(1)} + d^{(2)}\right)^2}{2N(2\hat{r}+1)}, \hat{\sigma}^2 = \hat{r}\hat{v}^2, \tag{7.12}$$

where $r = \hat{r}$ is the positive zero (provided it exists) of the quadratic equation

$$2\left(d^{(1)} - d^{(2)}\right)^2 r^2 - \left[\left(d^{(1)} + d^{(2)}\right)^2 - 2\left(d^{(1)} - d^{(2)}\right)^2\right] r - 2d^{(1)T}d^{(2)} = 0. \tag{7.13}$$

It may be noted that these estimates are independent of the location parameter estimate $\hat{\kappa}$. ■

The parameters $\left(\chi_i^{(n)} - \chi_1^{(n)}\right)$ and $\left(u + \chi_1^{(n)} + \rho_j^{(n)}\right)$ were estimated by means of (4.36), yielding the results displayed in Table 7-2. The "true values" are those underlying the data simulation, and are consistent with those appearing in Table 7-1. The differences of the estimated values from these is partly due to the noise inserted into the simulation.

The dispersion parameters $\sigma, v$ are estimated from (7.12) and (7.13), with the results appearing in Table 7-3. It is interesting to note that all of these results have been obtained in closed form, despite the relative complexity of the model.

### Table 7-2 Estimated location parameter values

| i or j | Data set n=1 | | | | Data set n=2 | | | |
|---|---|---|---|---|---|---|---|---|
| | $exp\left(\chi_i^{(n)} - \chi_1^{(n)}\right)$ | | $exp\left(u + \chi_1^{(n)} + \rho_j^{(n)}\right)$ | | $exp\left(\chi_i^{(n)} - \chi_1^{(n)}\right)$ | | $exp\left(u + \chi_1^{(n)} + \rho_j^{(n)}\right)$ | |
| | True value | Estimate | True value | Estimate | True value | Estimate | True value | Estimate |
| 1 | 1.000 | 1.000 | 232 | 248 | 1.0000 | 1.000 | 3485.5 | 3569.9 |
| 2 | 1.020 | 0.921 | 349 | 364 | 0.9802 | 0.896 | 13942.0 | 14398.8 |
| 3 | 1.041 | 0.922 | 581 | 636 | 0.9608 | 0.906 | 10456.5 | 10880.9 |
| 4 | 1.062 | 1.221 | 1162 | 1295 | 0.9418 | 0.959 | 5228.3 | 5441.2 |
| 5 | 1.083 | 1.060 | 1743 | 1899 | 0.9231 | 0.876 | 1045.7 | 1098.0 |
| 6 | 1.105 | 1.046 | 1743 | 1752 | 0.9048 | 0.883 | 348.6 | 350.1 |
| 7 | 1.127 | 1.081 | 1394 | 1511 | 0.8869 | 0.868 | 174.3 | 183.2 |
| 8 | 1.150 | 1.057 | 1162 | 1143 | 0.8694 | 0.851 | 87.1 | 86.6 |
| 9 | 1.174 | 0.963 | 929 | 848 | 0.8521 | 0.789 | 34.9 | 35.5 |
| 10 | 1.197 | 1.159 | 813 | 836 | 0.8353 | 0.878 | 17.4 | 18.8 |
| 11 | 1.221 | 1.107 | 581 | 591 | 0.8187 | 0.804 | 7.0 | 7.2 |
| 12 | 1.246 | 1.050 | 465 | 508 | 0.8025 | 0.869 | 7.0 | 7.8 |
| 13 | 1.271 | 1.338 | 290 | 285 | 0.7866 | 0.785 | 7.0 | 7.1 |
| 14 | 1.297 | 1.347 | 116 | 106 | 0.7711 | 0.805 | 7.0 | 6.8 |
| 15 | 1.323 | 1.334 | 58 | 52 | 0.7558 | 0.813 | 7.0 | 6.8 |

**Table 7-3 Estimated dispersion parameter values**

| Parameter | True value | Estimate |
|---|---|---|
| $\sigma$ | 0.100 | 0.088 |
| $\nu$ | 0.150 | 0.124 |

Statistics relating to the estimated dependency between the two data sets are also extracted. In particular,

- $corr(d^{(1)}, d^{(2)}) = 0.36$;
- Number of pairs $(i,j)$ for which $sgn\ d_{ij}^{(1)} = sgn\ d_{ij}^{(2)}$ is 144 out of 225 (64%).

Forecasting is a fairly straightforward application of Section 4.3, though it is worth noting the precise form of $\Omega^*$, defined by (4.43). Substitution of (7.2) into this definition gives

$$\Omega^* = M^*(M^T\Sigma^{-1}M)^{-1}M^{*T} + \hat{\sigma}^2(1_N 1_N^T) \otimes I_{|\mathcal{A}^*|} + \hat{\nu}^2 I_{N|\mathcal{A}^*|}\ , \tag{7.14}$$

where $|\mathcal{A}^{(n)*}|$ has been written as just $|\mathcal{A}^*|$ .

the results of the forecasts appear in Table 7-4, where forecasts are displayed for each of Data sets 1 and 2, and for the aggregation of the two. The CoV of the aggregate loss reserve is estimated to be 10.7%, and it is noteworthy that it would have been 8.5% if the forecasts in respect of the separate data sets were independent.

This amounts to a correlation of 58% between the two component loss reserves, causing a 25% increase in the standard error of the aggregate reserve. this increase would translate into a substantial increase in any VaR-based risk margin.

**Table 7-4 Forecast loss reserves**

| | Data set | | |
|---|---|---|---|
| | **1** | **2** | **1 and 2** |
| Forecast | 85953 | 60628 | 146581 |
| Standard error | 9781 | 7771 | 15639 |
| Coefficient of variation | 11.4% | 12.8% | 10.7% |

# 8. Conclusion

Section 4 formulates a model of multiple claim triangles subject to one of more common shocks. All observations are assumed to be log normally distributed, and cells means are log-linear.

The cell mean structure is very general. It consists of an idiosyncratic term (not part of the common shock structure) and, optionally, an "across-triangles" common shock and "within-triangle" common shocks. The idiosyncratic term is defined by a linear response that is arbitrary. As the names suggest, an across-triangles common shock applies a linear transformation of the same shocks to cells within different triangles; a within-triangle common shock applies a linear transformation of the same shocks to cells within the one triangle. In view of the log-linear structure of cell means, all shocks act multiplicatively, solving the problem of balanced shocks found in earlier papers.

The arbitrariness of the idiosyncratic log-linear term enables the accommodation of a very wide range of models. noteworthy is that the chain ladder is among them, but much more complex models can be accommodated.

The log-linear structure, coupled with log normally distributed observations, renders parameter estimation a mere matter of linear regression, from which closed form solutions may be obtained if the relevant variance matrices are known. Even when they too require estimation, it can sometimes be achieved in closed form, as in the example of Section 7.

The normal distribution is a member of the Tweedie family. As a result, the log normal distribution of individual cells may be generalized to log Tweedie, while the whole of the remainder of the model structure is retained (Section 5). Of course, this is accompanied by the loss of the simplicity of linear regression. A GLM is required instead, and closed form solutions are not available in most cases.

Perhaps the greatest reservation about the model presented here relates to the suitability of the log normal assumption for the aggregate distributions of individual cells. It may be noted that, if each cell can be regarded as represented by a compound distribution with sub-exponential severity tail (often realistic), then the compound distribution will also have a sub-exponential tail. This is likely to render the log Tweedie family (and possibly log normal) suitable for the aggregate distribution.

However, care is required in application of this assumption. For example, if the claim experience is large, with a large expected cell frequency, the aggregate distribution may appear close to normal, even though the tails are technically still sub-exponential, and one may asses the log normal assumption as unsuitable. Suffice to say that it suitability from one data set to another will depend on the detail of the data.

# Appendix A

**Maximum likelihood estimation for the model of Example 4.8**

Consider the model (4.31) and (4.32) with variance structure given by (4.28). As already noted, the vector of parameters defining this structure is $\omega = (\omega_1, \ldots, \omega_{2N+1})$, with components given by $\sigma^2, \left(\tau^{(1)}\right)^2, \ldots, \left(\tau^{(N)}\right)^2, \left(v^{(1)}\right)^2, \ldots, \left(v^{(N)}\right)^2$, and the problem is to locate the roots $\omega_k$ of (4.40).

The derivatives of $\Sigma$ in (4.40) are obtained from (4.30), as follows:

$$\frac{\partial \Sigma}{\partial \sigma^2} = (1_N 1_N^T) \otimes \left(A^{(0)} R \left(A^{(0)}\right)^T\right), \quad (A.1)$$

$$\frac{\partial \Sigma}{\partial (\tau^{(n)})^2} = (e_n e_n^T) \otimes \left(A^{(0)} R \left(A^{(0)}\right)^T\right), \quad (A.2)$$

$$\frac{\partial \Sigma}{\partial (v^{(n)})^2} = (e_n e_n^T) \otimes I_{|\mathcal{A}|}, \quad (A.3)$$

where $e_n$ denotes the $n$-th natural co-ordinate (column) vector, consisting of all zeros, except unity in the $n$-th position.

# Appendix B

**Derivation of Proposition 7.1**

Consider the calculation of the quantities $\partial\ell/\partial\sigma^2, \partial\ell/\partial v^2$ according to (4.40). One can check that

$$\Sigma^{-1}\frac{\partial \Sigma}{\partial \sigma^2} = \frac{1}{(2\sigma^2 + v^2)}\begin{bmatrix} I & I \\ I & I \end{bmatrix}, \Sigma^{-1}\frac{\partial \Sigma}{\partial v^2} = \Sigma^{-1} = \frac{1}{v^2(2\sigma^2 + v^2)}\begin{bmatrix} (\sigma^2 + v^2)I & -\sigma^2 I \\ -\sigma^2 I & (\sigma^2 + v^2)I \end{bmatrix}, \quad (B.1)$$

$$tr\left(\Sigma^{-1}\frac{\partial \Sigma}{\partial \sigma^2}\right) = \frac{2N}{2\sigma^2 + v^2}, tr\left(\Sigma^{-1}\frac{\partial \Sigma}{\partial v^2}\right) = \frac{2N(\sigma^2 + v^2)}{v^2(2\sigma^2 + v^2)}, \quad (B.2)$$

$$\Sigma^{-1}\frac{\partial \Sigma}{\partial \sigma^2}\Sigma^{-1} = \frac{1}{(2\sigma^2 + v^2)^2}\begin{bmatrix} I & I \\ I & I \end{bmatrix}, \Sigma^{-1}\frac{\partial \Sigma}{\partial v^2}\Sigma^{-1} = \frac{1}{v^4(2\sigma^2 + v^2)^2}\begin{bmatrix} ((\sigma^2 + v^2)^2 + \sigma^4)I & -2\sigma^2(\sigma^2 + v^2)I \\ -2\sigma^2(\sigma^2 + v^2)I & ((\sigma^2 + v^2)^2 + \sigma^4)I \end{bmatrix}. \quad (B.3)$$

Let $d$ denote the vector $y - \hat{y}$, and dissect $d$ as follows: $d^T = \left(d^{(1)T}, d^{(2)T}\right)$, where $d^{(n)}$ relates to data set $n$. Substitution of (B. 1)-(B. 3) into (4.40) yields, after a little manipulation,

$$\left(d^{(1)} + d^{(2)}\right)^2 = 2N(2\sigma^2 + v^2), \quad (B.4)$$

$$((\sigma^2 + v^2)^2 + \sigma^4)\left(d^{(1)} - d^{(2)}\right)^2 + 2v^4 d^{(1)T} d^{(2)} = 2Nv^2(\sigma^2 + v^2)(2\sigma^2 + v^2). \quad (B.5)$$

Set $r = \sigma^2/v^2$ and divide the last equation through by $v^4$ to convert it to the form

$$((1+r)^2+r^2)\left(d^{(1)}-d^{(2)}\right)^2+2d^{(1)T}d^{(2)}=2N(1+r)(2\sigma^2+v^2), \tag{B. 6}$$

which, by (B. 4), may be expressed as

$$((1+r)^2+r^2)\left(d^{(1)}-d^{(2)}\right)^2+2d^{(1)T}d^{(2)}=(1+r)\left(d^{(1)}+d^{(2)}\right)^2. \tag{B. 7}$$

The quantities $d^{(1)}, d^{(2)}$ are known, and so this is a solvable quadratic in $r$. With an estimate of $r$, it is now possible to estimate values of $\sigma^2, v^2$. From (B. 4),

$$\left(d^{(1)}+d^{(2)}\right)^2=2Nv^2(2r+1), \tag{B. 8}$$

and the estimate of $v^2$ stated in the proposition follows. ■

# Appendix C

## Synthetic data rectangles

### Data set 1

| Accident year | Claim payments in development year | | | | | | | | | | | | | | |
|---|---|---|---|---|---|---|---|---|---|---|---|---|---|---|---|
| | 1 | 2 | 3 | 4 | 5 | 6 | 7 | 8 | 9 | 10 | 11 | 12 | 13 | 14 | 15 |
| 1 | 298 | 336 | 569 | 1153 | 1537 | 1669 | 1730 | 990 | 1096 | 1091 | 548 | 472 | 261 | 118 | 52 |
| 2 | 222 | 478 | 558 | 787 | 1925 | 1708 | 1124 | 1448 | 665 | 824 | 512 | 412 | 344 | 88 | 62 |
| 3 | 258 | 331 | 651 | 1029 | 1505 | 1775 | 1916 | 741 | 695 | 709 | 682 | 567 | 218 | 147 | 65 |
| 4 | 320 | 438 | 626 | 1698 | 2982 | 2071 | 1442 | 1565 | 1262 | 1111 | 547 | 628 | 344 | 126 | 64 |
| 5 | 221 | 317 | 656 | 1542 | 1881 | 2186 | 1885 | 1230 | 970 | 688 | 758 | 578 | 262 | 128 | 67 |
| 6 | 401 | 542 | 527 | 943 | 1786 | 1736 | 1443 | 1336 | 919 | 804 | 730 | 958 | 260 | 113 | 59 |
| 7 | 215 | 504 | 673 | 1906 | 1982 | 1831 | 1911 | 1099 | 688 | 818 | 526 | 551 | 303 | 151 | 77 |
| 8 | 251 | 296 | 704 | 1626 | 2083 | 1942 | 1529 | 1274 | 1175 | 1027 | 574 | 485 | 384 | 166 | 56 |
| 9 | 227 | 326 | 775 | 1188 | 2444 | 1420 | 1213 | 1421 | 1039 | 1167 | 586 | 778 | 345 | 171 | 68 |
| 10 | 303 | 385 | 852 | 1600 | 1894 | 1991 | 1830 | 1537 | 770 | 1470 | 677 | 647 | 285 | 127 | 57 |
| 11 | 173 | 346 | 797 | 2232 | 2208 | 2234 | 1419 | 1758 | 1007 | 911 | 482 | 577 | 531 | 194 | 61 |
| 12 | 216 | 457 | 734 | 1251 | 2610 | 1868 | 1469 | 1597 | 995 | 1309 | 514 | 533 | 369 | 85 | 96 |
| 13 | 425 | 414 | 783 | 1509 | 1694 | 2047 | 1367 | 1476 | 1006 | 1080 | 875 | 771 | 388 | 140 | 75 |
| 14 | 365 | 450 | 902 | 1564 | 2486 | 1534 | 1878 | 1395 | 1178 | 1033 | 756 | 464 | 407 | 175 | 73 |
| 15 | 331 | 402 | 684 | 1625 | 2843 | 2689 | 1669 | 1389 | 1158 | 937 | 648 | 573 | 363 | 171 | 54 |

### Data set 2

| Accident year | Claim payments in development year | | | | | | | | | | | | | | |
|---|---|---|---|---|---|---|---|---|---|---|---|---|---|---|---|
| | 1 | 2 | 3 | 4 | 5 | 6 | 7 | 8 | 9 | 10 | 11 | 12 | 13 | 14 | 15 |
| 1 | 3062 | 14153 | 10645 | 5835 | 997 | 287 | 208 | 77 | 41 | 19 | 8 | 8 | 7 | 8 | 7 |
| 2 | 2821 | 14536 | 9553 | 4506 | 1238 | 272 | 154 | 82 | 34 | 16 | 6 | 8 | 7 | 5 | 6 |
| 3 | 4187 | 11794 | 9420 | 4520 | 999 | 357 | 153 | 82 | 32 | 15 | 8 | 7 | 5 | 7 | 7 |
| 4 | 4008 | 12451 | 10124 | 5437 | 1013 | 344 | 126 | 88 | 46 | 20 | 6 | 7 | 6 | 5 | 9 |
| 5 | 3601 | 10050 | 10720 | 4762 | 908 | 330 | 200 | 73 | 25 | 17 | 6 | 6 | 7 | 7 | 7 |
| 6 | 3373 | 14509 | 9924 | 3564 | 950 | 351 | 157 | 94 | 26 | 16 | 7 | 7 | 6 | 5 | 6 |
| 7 | 3578 | 13356 | 7416 | 4644 | 1031 | 298 | 173 | 76 | 28 | 18 | 7 | 6 | 7 | 7 | 6 |
| 8 | 2875 | 13080 | 9625 | 4913 | 938 | 328 | 157 | 59 | 33 | 16 | 5 | 6 | 7 | 7 | 7 |
| 9 | 2930 | 13345 | 7174 | 4189 | 833 | 268 | 155 | 75 | 29 | 18 | 6 | 6 | 6 | 6 | 7 |
| 10 | 2558 | 12815 | 11495 | 5197 | 935 | 293 | 112 | 82 | 30 | 16 | 5 | 6 | 7 | 5 | 6 |
| 11 | 2366 | 12672 | 8330 | 5280 | 852 | 329 | 139 | 71 | 33 | 14 | 5 | 6 | 6 | 6 | 6 |
| 12 | 2617 | 12305 | 10697 | 5039 | 894 | 203 | 138 | 62 | 23 | 15 | 4 | 5 | 6 | 6 | 6 |
| 13 | 2984 | 9702 | 9356 | 3508 | 790 | 294 | 115 | 84 | 34 | 14 | 6 | 7 | 6 | 5 | 5 |
| 14 | 2963 | 11253 | 10616 | 3545 | 713 | 230 | 141 | 57 | 29 | 12 | 6 | 6 | 6 | 5 | 5 |
| 15 | 2902 | 10489 | 8273 | 3259 | 768 | 358 | 108 | 55 | 27 | 13 | 5 | 5 | 5 | 5 | 5 |